\documentclass[12pt,preprint]{emulateapj}
\usepackage{color}
\usepackage{natbib}
\shorttitle{A plausible (overlooked) super-luminous supernova in the SDSS Stripe 82 data}
\shortauthors{Kostrzewa-Rutkowska et al.}

\begin{document}

\title{A plausible (overlooked) super-luminous supernova in the SDSS Stripe 82 data}

\author{Zuzanna Kostrzewa-Rutkowska\altaffilmark{1}, 
Szymon Koz{\l}owski\altaffilmark{1}, 
{\L}ukasz Wyrzykowski\altaffilmark{1,2}, 
S. George Djorgovski\altaffilmark{3},
Eilat~Glikman\altaffilmark{4},
Ashish A. Mahabal\altaffilmark{3},
Sergey Koposov\altaffilmark{2}
}

\altaffiltext{1}{Warsaw University Observatory, Al.\ Ujazdowskie 4, 00-478 Warszawa, Poland, email: \{zkostrzewa, simkoz, wyrzykow\}@astrouw.edu.pl}
\altaffiltext{2}{Institute of Astronomy, University of Cambridge, Madingley Road, Cambridge CB3 0HA, UK}
\altaffiltext{3}{California Institute of Technology, 1200 E California Blvd., Pasadena, CA 91125, USA}
\altaffiltext{4}{Department of Physics and Yale Center for Astronomy and Astrophysics, Yale University, P.O. Box 208121, New Haven, CT 06520-8121, USA}

\begin{abstract}
We present the discovery of a plausible super-luminous supernova (SLSN), 
found in the archival data of Sloan Digital Sky Survey (SDSS) Stripe 82, called 
PSN~000123+000504.
The supernova peaked at $M_{\rm g}<-21.3$~mag
 in the second half of September 2005, but was missed by the real-time supernova hunt.
 The observed part of the light curve (17 epochs) showed that 
the rise to the maximum took over 30 days, while the decline time lasted at least 70 days (observed frame), closely resembling other SLSNe of SN2007bi type.
Spectrum of the host galaxy reveals a redshift of $z=0.281$ 
and the distance modulus of $\mu=40.77$~mag. 
Combining this information with the SDSS 
photometry, we found the host galaxy to be an LMC-like irregular dwarf galaxy with the 
absolute magnitude of $M_B=-18.2\pm0.2$~mag and the oxygen abundance of 
${\rm 12+\log[O/H]}=8.3\pm0.2$. 
Our SLSN follows the relation for the most energetic/super-luminous 
SNe exploding in low-metallicity environments, but we found no clear evidence for SLSNe to explode in low-luminosity (dwarf) galaxies only. 
The available information on the PSN~000123+000504 light curve suggests the magnetar-powered model as a likely scenario of this event.
This SLSN is a new addition to a quickly growing family of
super-luminous SNe.
\end{abstract}


\section{Introduction}

There is a growing number of supernovae (SNe) whose brightness greatly exceeds that of ``classic'' 
SNe (e.g. \citealp{2011Natur.474..487Q}), hence not fitting into any of the standard SNe classes. 
It is widely assumed, albeit somewhat arbitrarily, that an SN becomes an SLSN if its absolute magnitude 
exceeds $-21.0$~mag in any filter, what corresponds to the peak luminosity of $>7\times10^{43}$~erg/s. 
During the last few years most of SLSN events have been discovered by several untargeted surveys, for example the Catalina Real-Time Transient 
Survey (\citealp{2009ApJ...696..870D}) and the Palomar Transient Factory (\citealp{2009PASP..121.1395L}; 
\citealp{2009PASP..121.1334R}).
It was found that these over-luminous SNe tend to prefer 
low-metallicity, low-mass dwarf galaxies  (\citealp{2008ApJ...673..999P,2006AcA....56..333S,2011ApJ...727...15N}).

The SLSNe are associated with deaths of the most massive stars, which means that they have impact on the chemical 
evolution and re-ionization of the Universe. 
The SLSNe explosions are probably induced by various physical mechanisms than other, more common types of SNe, 
which makes these transients more interesting. 
However, still little is known about the nature of SLSNe, because of 
insufficient size and low heterogeneity of the available sample (for a recent review, see \citealp{2012Sci...337..927G}).
The SLSNe family has been divided into three classes: SLSN-I (hydrogen-poor), SLSN-II (hydrogen-rich), and SLSN-R (radioactively powered).
SLSN-R type seems to be the best understood one as these SNe are powered by large amounts of radioactive $^{56}$Ni produced during the explosions of massive stars.
The mechanism of the most commonly observed SLSN-II class can be explained by an explosion within a thick hydrogen envelope, but the proper energy source is still unknown.
The final group, SLSN-I explosions, is probably a result of pair-instability mechanism, but these SNe are not powered by radioactivity.
Recently a different statement was presented by \citet{2013arXiv1304.3320I}. 
These authors claim that type SLSN-I can be classified as SLSN-Ic, because their spectral features are similar to typical Ic explosions, i.e.\ the blue continuum with broad absorption lines of intermediate mass elements such as C, O, Si, and Mg, with no clear evidence of H and He.
Moreover, the energy deposited by newborn magnetars was favored as the power source for these events instead of the pair-instability mechanism.
A semi-analytical diffusion model with energy input from the spin-down of a magnetar reproduces the extensive light curve data well. 
The ejecta velocities and temperatures required by the model predictions are in a reasonable agreement with those determined from the photometric and spectroscopic observations. 
There are therefore some discrepancies in views on the real nature of the SLSNe, which is caused by a very small number of known examples of such explosions. 

In this paper, we present the discovery of another object belonging to the class of SLSNe found in the SDSS Stripe 82 multi-band data. 
To date there are $\sim$30 known SLSNe, therefore each new event observed is a valuable addition to the class.
With its peak brightness of $M_g<-21.3$~mag (without including the $K$-correction and host galaxy extinction) 
and the optical light curve shape, it closely resembles SLSNe.
The subsequent sections include a detailed description of our findings.


\section{The Supernova}
\label{sec:SN}

\begin{figure}
\centering
\includegraphics[width=8cm]{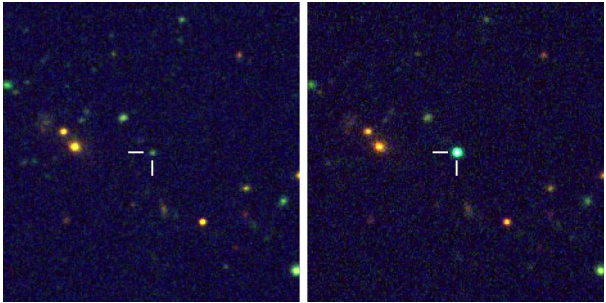}
\caption{Finding chart for PSN~000123+000504. In the left panel, cross-hair marks the SN host galaxy, while the right panel shows the image taken  near the peak of the SN explosion.
The RGB image is a composite of $u$, $r$, and $i$-band filters and covers $1.5'\times 1.5'$. North is up, east is to the left.}
\label{fig:finding_chart}
\end{figure}

\begin{figure}
\centering
\includegraphics[width=8cm]{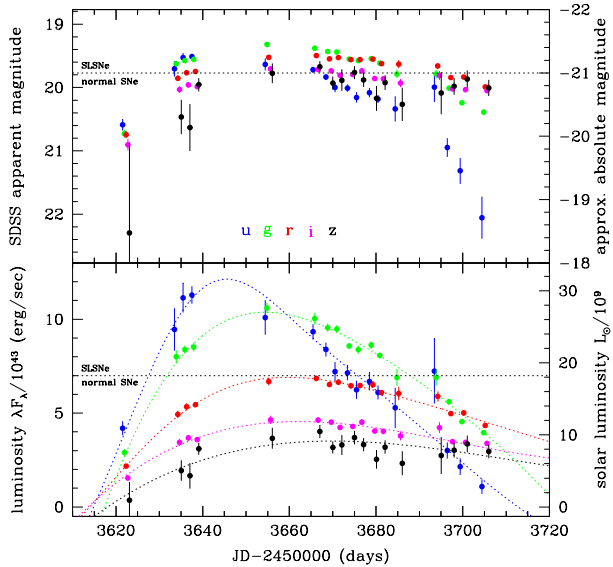}
\caption{Light curves of PSN~000123+000504. {\it Top panel:} The SN extinction-corrected observed magnitudes
in the AB magnitude system and the approximate absolute magnitudes (host extinction and $K$-corrections not included) 
for five SDSS filters (marked with colors). For clarity, data points are shifted by $-0.8$ ($u$-band), $-0.4$ ($g$-band), $+0.4$ ($i$-band), $+0.8$ ($z$-band) days with respect to the observed date.
The horizontal dashed line marks the division between normal SNe and SLSNe from \cite{2012Sci...337..927G}.
{\it Bottom panel:} The absolute magnitudes from the top panel converted to luminosity.
The SN peaked above $10^{44}$~erg/s, typical for SLSNe and atypical for normal SNe. 
The combined peak luminosity from five filters exceeded $3.5\times10^{44}$~erg/s (Table~\ref{tab:luminosity}). }
\label{fig:lc}
\end{figure}

The Sloan Digital Sky Survey (SDSS) repeatedly observed several selected areas of the sky.
One of them, called Stripe 82, covering $\sim$270 deg$^2$ of the sky, was  specifically targeted for multiple photometric observations (\citealt{2007AJ....134..973I}) in the hunt for type Ia supernovae explosions.
There were about 500 SNe found in near-real-time data analysis, many of them followed-up spectroscopically (\citealt{2008AJ....135..338F, 2008AJ....135..348S}).
The Stripe~82 covers the area of $-50^\circ<\alpha<59^\circ$, $-1.25^\circ<\delta<1.25^\circ$.
We used the catalog of \citet{2008MNRAS.386..887B} that provided not only the photometric and astrometric time-series for almost 4 million objects,
 but also numerous high-level data products, useful for an effective querying of the catalog. 
We aimed at transients which lasted between 40 and 150 days and had mean luminosities
brighter than 20.5 mag in $g$ and $u$ bands.
In addition, we only looked for objects with a significant excess in blue at the peak.
We found about 200 potentially interesting objects, most of them turning out to be previously found, known SNe (\citealt{2008AJ....135..338F}, \citealt{2008AJ....135..348S}).
From the remainder we selected one transient with well covered light curve, exhibiting long rise and long decline, as a possible SLSN. 
The transient was also located on top of a faint galaxy-like object (according to the SDSS Data Release 9; DR9\footnote{http://skyserver.sdss3.org/public/en/tools/chart/navi.asp}), which ruled out the possibility for a cataclysmic variable transient.
The coordinates of the SN are $(\alpha,\delta)$=(00:01:23.36,+00:05:04.02) and the finding chart is presented in Figure \ref{fig:finding_chart}.
We dubbed the supernova with PSN~000123+000504 aka ``Vernal Equinox Supernova 2005''  because of its position very close to the beginning of the equatorial coordinate system.
The observed light curve of the SN is shown in Figure~\ref{fig:lc}. 
It is also available in tabular form for the
observed magnitudes (Table~\ref{tab:observed}), absolute magnitudes (Table~\ref{tab:absolute}), and luminosity (Table~\ref{tab:luminosity}).

It is not clear why the object was not found by a regular search for SNe during the Stripe 82 observations, which were designed for real-time supernova hunt.
Most probably, the host galaxy was not recognized as a galaxy due to its compactness and blue color, suggesting a galactic progenitor of some sort of a cataclysmic variable.
Only after producing the final SDSS images of this region (in SDSS DR9) the object was classified as a galaxy. 

While we do not have the spectrum for the SN during explosion, we can only rely on post-explosion observations. 
In the next section we present the evidence for the SN host galaxy redshift of $z=0.281$ 
which corresponds to the distance modulus of $\mu=40.77$~mag and the luminosity distance of 1428 Mpc,
assuming a standard $\Lambda$CDM cosmological model with 
$(H_0, \Omega_M, \Omega_{\rm vac}, \Omega_k)=(71~{\rm{km/s/Mpc}}, 0.27, 0.73, 0.0)$.
The Milky Way extinctions in the direction of the SN are 
$(A_u, A_g, A_r, A_i, A_z)=(0.128, 0.100, 0.069, 0.051, 0.038)$~mag 
(\citealt{1998ApJ...500..525S}).

The absolute magnitude of the SN observed at magnitude $m$ is described by the equation:
\begin{equation} 
\label{eq:abs}
M = m - \mu - A_{\rm MW} - A_{\rm H}(z) - K(z,~A_{\rm H}(z)), 
\end{equation} 
where $\mu=5\log{d(z, \Omega_X)}+25$ is the distance modulus, $d(z, \Omega_X)$ is the luminosity distance in Mpc, a function of redshift $z$ and 
cosmological model $\Omega_X$, $A_{\rm MW}$ and $A_{\rm H}(z)$ are the extinctions in our own 
Milky Way galaxy and in the SN host galaxy, respectively, and $K(z,~A_{\rm H}(z))$ is the single 
filter {\it K}-correction (see e.g., \citealt{1996PASP..108..190K}).

There are two unknowns in Equation~\ref{eq:abs}. First of all, we do not know the value of
$A_{\rm H}(z)$, that is the internal host extinction at the SN location. 
However,  because it can only make the SN brighter (the larger the value the brighter the SN), it does not affect our interpretation of the SN as super-luminous one.
The second unknown is the {\it K}-correction, as for its calculation we would need a spectrum near the peak.
On the other hand, the {\it K}-correction at such a low redshift is not going to change the absolute magnitudes by more than a few tenths of magnitude  in either direction.

\begin{figure}
\centering
\includegraphics[width=8cm]{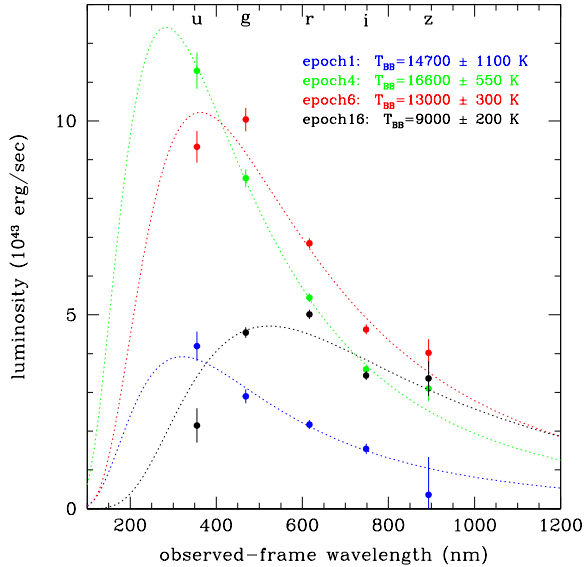}
\caption{Black-body at $z=0.281$ fits to the SED constructed from the photometric multi-band data for epochs 1 (blue), 4 (green), 6 (red), and 16 (black), corresponding to $-16$, $0$, $+28$, and $+62$~days from the $u$-band peak.
While the fits to the pre-peak epochs return good black-body matches (with $\chi^2/{\rm dof}<1.5$), 
this is not the case for late epochs. This is indicative of a significant line evolution at late epochs, 
characteristic for core-collapse SNe. The luminosities were corrected for the Galactic extinction.}
\label{fig:temp_prof}
\end{figure}

We converted the absolute magnitudes to luminosity (bottom panel of Figure~\ref{fig:lc}, Figure~\ref{fig:temp_prof}, and Table~\ref{tab:luminosity}).
Integrating the five-band SDSS light curves over the duration of the transient, we estimated that the total emitted energy exceeded  $2.1\times10^{51}$ ergs. 
The bottom panel of Figure~\ref{fig:lc} shows the luminosity light curves. 
The first obvious finding is that the $u$-band (near UV) light curve peaks first (at epoch 4 of the light curve),
followed by optical light curves (at epochs 5--6), while the infrared light curves peak as the last ones, at more later epochs. 
This obviously reflects the changes in the spectrum of the object that are characteristic for an object with high early temperatures and low temperatures at late times -- clear signs of an explosion or eruption.
For each epoch we constructed five-point spectral energy distribution (SED) from the photometric data and investigated the evolution of the SED shape with  time.
In Figure~\ref{fig:temp_prof}, we present the supernova SED at selected epochs 1, 4, 6, and 16, corresponding to $-16$, $0$, $+28$, and $+62$~days from the $u$-band peak at  ${\rm JD}=2453638.32$ (24 September 2005).
We fitted a black-body (BB) spectrum at each epoch and obtained the best-fitting BB temperatures ($T_{{\rm BB}}$). 
The BB fits to the early time (pre-peak) SEDs returned satisfying $\chi^2/{\rm dof}<1.5$, pointing to a rather featureless spectra with high temperatures $T_{{\rm BB}}\gtrsim15000$~K.
The situation changes when we fit the BB spectrum to the late time (post-peak) SEDs. 
The fits are statistically unacceptable with $7<\chi^2/{\rm dof}<14$. 
This is easily explained if we assume that our object is to an SN. 
\cite{1997ARA&A..35..309F} describes a normal SN Type II-P as having a featureless blue BB spectrum $T_{{\rm BB}}>10000$~K at early times
and a significant line evolution at late times with temperature reaching $\sim5000$~K few weeks later (adiabatic expansion and cooling of the ejecta) -- a picture perfectly fitting into our findings (if ignoring the high luminosity).

\section{The Host of PSN~000123+000504}
\label{sec:host}

The SN was found to be a transient in the Stripe 82 data, i.e.\ with no prior historical observations, forming a baseline. 
However, there is a faint galaxy-like object present in the SDSS images (\citealp{2012ApJS..203...21A}) at the location of PSN~000123+000504.
Its brightness in the SDSS DR9 is $(u, g, r, i, z)=(24.50\pm0.91, 23.12\pm0.21, 
22.06\pm0.14, 21.88\pm0.16, 22.43\pm0.73)$~mag in the AB magnitude system. 

\begin{figure}
\centering
\includegraphics[width=8cm]{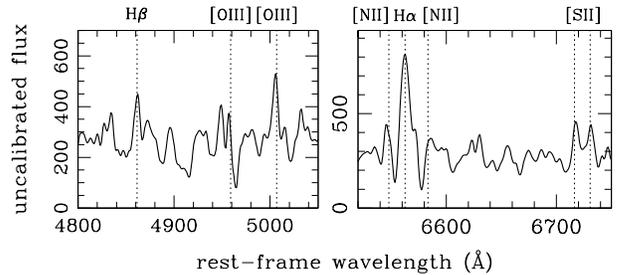}
\caption{Selected parts of the Keck/LRIS spectrum with the major host emission lines used for redshift determination.
The spectrum has a relatively low signal-to-noise ratio, nevertheless 
it was sufficient to secure the redshift of $z=0.281$ and host's metallicity of ${\rm 12+\log[O/H]}\approx8.3$.
Vertical dotted lines with labels mark major emission lines.}
\label{fig:keck_spec}
\end{figure}

We obtained a 30 min KECK/LRIS spectra of the host galaxy on MJD=56215.353347 and 56215.353455 (15 October 2012).
While the spectrum has a rather low overall signal-to-noise (S/N) ratio, we were able to identify (amongst other) the following major emission lines: 
H$\beta$ at 4861\AA,
[OIII] at 4959\AA\ and 5007\AA,
H$\alpha$ at 6863\AA,
and [NII] at 6884\AA, at a redshift $z=0.281$ (Figure \ref{fig:keck_spec}). 

\begin{figure}
\centering
\includegraphics[width=8cm]{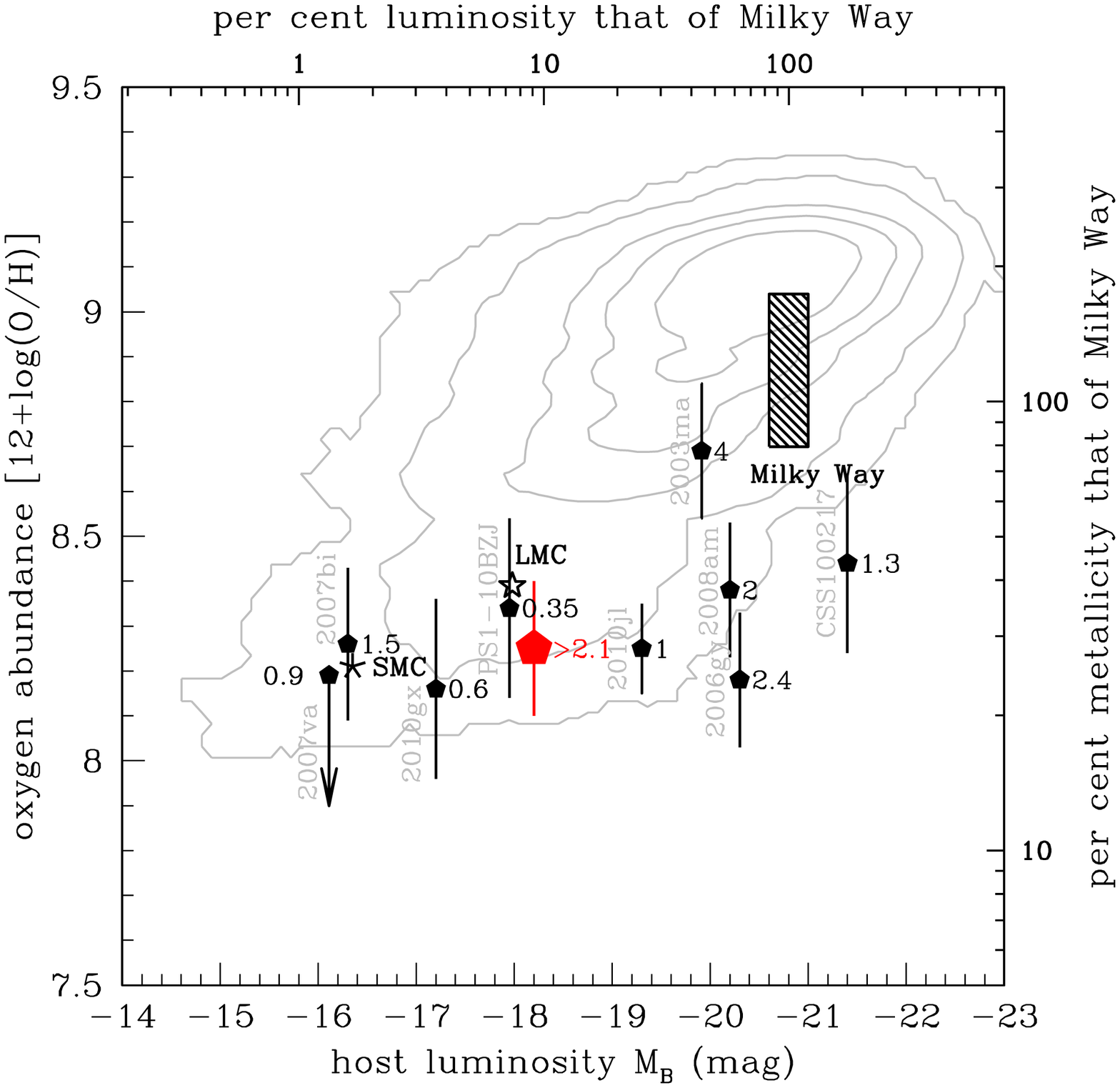}
\caption{Metallicity--luminosity plane for 
125,958 SDSS DR4 galaxies from \citealt{2008ApJ...673..999P} on the metallicity scale from \cite{2004ApJ...613..898T}. 
The plane was divided into $\Delta M_{\rm B}=0.1$ mag and
$\Delta ({\rm 12+\log[O/H]})=0.025$~dex bins. 
The contours are for 2, 10, 50, 100, 
and 200 galaxies per bin, smoothed with $3\times3$ bins window.
The Milky Way's, LMC's and SMC's locations are marked with a box, open star, and narrow star, respectively. 
Full pentagons mark nine SLSN hosts.
Numbers indicate the amount of energy 
(in units of $10^{51}$ ergs) emitted during these explosions 
(points taken from \citealt{2010ApJ...722.1624K,2011ApJ...730...34S}). 
The PSN~000123+000504 host galaxy is marked with a large red pentagon.}
\label{fig:oxy}
\end{figure}

The red part of the spectrum was then also used to estimate the host's oxygen abundance (${\rm 12+\log[O/H]}$).
We measured the emission lines 
H$\beta$ with S/N $=7.4$,
[OIII] at 4959\AA\ with S/N $\approx0.5$,
[OIII] at 5007\AA\ with S/N $=11.3$,
H$\alpha$ with S/N $=19.7$,
and [NII] at 6884\AA\ with S/N $=2.8$
to measure the $N2$ and $O3N2$ metallicity indices. 
Using equations from
\cite{2004MNRAS.348L..59P} we found ${\rm 12+\log[O/H]}=8.29$ with a formal error
of 0.15 for $N2$ index, and ${\rm 12+\log[O/H]}=8.33$ with a formal error
of 0.08 for $O3N2$ index. 
We added in quadrature the uncertainty for the 
\cite{2004MNRAS.348L..59P} relations ($\sigma=0.19$ and 0.13, respectively)
to obtain ${\rm 12+\log[O/H]}=8.33\pm 0.15$ for $O3N2$ index and
${\rm 12+\log[O/H]}=8.29\pm0.24$ for $N2$ index (Figure~\ref{fig:oxy}). 
Both values are consistent with each other and indicate that the SN host galaxy has
30\% Milky Way's metallicity (adopting 8.8 from \citealt{2010arXiv1005.0423D}).

After converting the host magnitudes to luminosity, we fit the galaxy/AGN models
of \cite{2010ApJ...713..970A} to the five photometric data points. 
The only well-fitting spectrum is that of an irregular galaxy. 
This supports our claim that the host galaxy is a dwarf, irregular, metal-poor galaxy.
The absolute brightness of the galaxy in $B$ was obtained as $M_{\rm B}=-18.2\pm0.2$ mag and the galaxy was placed on the absolute magnitude--metallicity plane, as shown in Figure~\ref{fig:oxy}.

\section{Discussion}
\label{sec:discussion}

In this paper, we presented the discovery of an overlooked supernova in the SDSS Stripe~82 archival data, 
which we classified as an SLSN one based on its peak optical magnitude brighter than $-21$.
The SN was found in a search for long transient events with a significant blue excess at peak in a variability database for SDSS Stripe~82.
While we do not have the spectrum for the SN during explosion, we present the evidence for the SN host galaxy redshift of $z=0.281$ using the KECK/LRIS spectrum.
We also constructed the five-point SED from the photometric data and investigated its evolution as a function of time.
We fitted the BB spectrum at each epoch and obtained a best-fitting BB temperature, greater than 15000K, in early time SEDs.
The temperature evolution with a decline in late epochs to $\sim$5000K also constitutes a strong piece of evidence in favor of classifying this object as a SLSN (\citealp{2009ApJ...690.1303M}; \citealp{2009ApJ...702..226M}).

\begin{figure}
\centering
\includegraphics[width=8cm]{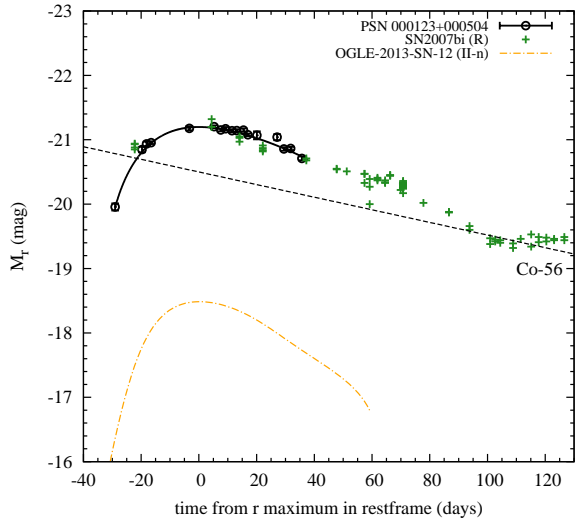}
\caption{The absolute $r$-band light curve of PSN~000123+000504 (``Vernal Equinox Supernova 2005'', black points) and another super-luminous event SN2007bi (\citealt{2009Natur.462..624G}). 
The light curve of SN2007bi is shown in the R-band. Also an SN Type II-n from the OGLE database is shown (OGLE-2013-SN-12, \citealp{2013ATel.4774....1E}). The solid black line is the spline fit to the SN data. The data for each SN have been corrected for time dilation and distance modulus.}
\label{fig:sne_lc}
\end{figure}

In Figure \ref{fig:sne_lc}, we present the $r$-band light curves of our SN.
For comparison we also show the light curve of a likely Type II-n SN from the OGLE database 
(\citealt{2013ATel.4774....1E}; \citealt{2012ATel.4495....1W}; see also \citealt{2013AcA....63....1K} for OGLE SNe overview), which is a few magnitudes fainter, but presents a similar shape of the light curve. 
Overall, the shape of the light curve of PSN~000123+000504 resembles very much that of objects found previously and classified as SLSNe. 
The rise of the light curve is very slow - it lasts about 30 days in the restframe from the faintest detected data point to the maximum.
Then, the decline is also very slow, in a rough agreement with a radioactive decay of Co-56, at least in its early post-peak phase. 
Such behavior was in the past associated with supernovae of SLSN-R type \citep{2012Sci...337..927G}, however, the suggested pair-instability origin of those SLSNe was recently questioned by 
\cite{2013arXiv1304.3320I}.
They suggest that SNe similar to SN2007bi are just a subtype of SLSN-I and are well described by a magnetar scenario.

There is a growing evidence for a correlation of the most energetic explosions  preferring to occur in the low-metallicity 
($<50$\% that of Milky Way) dwarf ($L/L_\star<0.1$) 
galaxies (\citealt{2006AcA....56..333S, 2010ApJ...722.1624K, 2011ApJ...730...34S, 2012Sci...337..927G}). 
While we find the former to be true for SLSN (Fig.~\ref{fig:oxy}), we are unable to confirm the latter part of the statement. 
In Fig.~\ref{fig:oxy}, we see that luminous SNe occupy all types of host galaxies with $-16<M_{\rm B}<-21.5$ mag (galaxies with 1 to 200 per cent MW luminosity). 
Hence, there is no clear evidence
for SLSNe to prefer dwarf galaxies, but this claim is based on a statistically small sample of 10 hosts.

We examined the properties of the faint host galaxy.
With an oxygen abundance $({\rm 12+\log[O/H]})\approx8.30$, the host galaxy of PSN~000123+000504 has 30\% of the Milky Way's metallicity 
and the absolute magnitude comparable to that of the Large Magellanic Cloud (LMC).
The place of the SN host galaxy on the absolute magnitude-metallicity plane (Fig.~\ref{fig:oxy}) is among other hosts with detected SLSNe.
The presence of SLSN usually indicates that the host is undergoing an episode of active star formation. 
The extreme luminosity of SLSNe is the key for their discovery over a range of redshifts, what helps the investigation 
of star formation in galaxies at high redshifts.
Furthermore, the plausible high masses of their progenitors (in excess of 150 M$_\odot$) present 
an opportunity to model the initial mass function in e.g., low-mass galaxies.
The low host masses of observed SLSNe would indicate that these objects were produced at the early stages of galactic evolution,
which creates a modeling problem to obtain progenitors in such galaxies with consistent metallicity (see \citealp{2010A&A...512A..70Y} for a comparison).

Based on our single detection we can roughly estimate the SLSNe rate from the SDSS data.
The Stripe 82 survey covers about 270 square degrees which corresponds to a volume of $\sim$$0.28~\mathrm{Gpc}^3$ within redshift $z<0.4$,
a distance to which we are able to detect our SLSN.
The brightest part of the SN light curve in range $-20$ to $-21.5$ mag can be seen during about 100 days (in $g$-band with a detection limit of 21.5 mag).
During the 8 year survey only one SLSN has been detected. 
Hence, we find that the SLSNe rate is of order of $-9 <\log(\mathrm{SLSN}~\mathrm{Mpc}^{-3} \mathrm{yr}^{-1}) <-8$ (with the assumption of 10 per cent efficiency in SN detection for Stripe 82 data, which is low mostly due to non-uniform SDSS cadence).
This result is consistent with the rates presented previously by \cite{2012Sci...337..927G}.

\section{Summary}
We presented a new example of the SLSNe class, which was found in the archival data of the SDSS Stripe 82. 
The redshift of the host, $z=0.281$, determined with the Keck/LRIS spectrum, indicates that the brightness of this SN in the maximum exceeded $-21$ mag.
The light curve resemblance to SN2007bi and other similar SLSNe discovered so far indicates the PSN~000123+000504 may belong to the SLSN-R class (\citealt{2009Natur.462..624G,2010A&A...512A..70Y}) or to magnetars (\citealt{2013arXiv1304.3320I}).
This study also shows the potential of archival multi-color time-series databases in successful detection of interesting and rare cases of transients. 


\acknowledgments

Based in part on the data obtained at the W.M. Keck Observatory, which is operated as a scientific partnership among the California Institute of Technology, the University of California and the National Aeronautics and Space Administration. The Observatory was made possible by the generous financial support of the W.M. Keck Foundation.

ZKR is supported from grant 2011/01/N/ST9/03069 by the Polish National Science Centre.
SK is supported from (FP7/2007-2013)/ERC grant agreement no.\ 246678 awarded by the European Research Council
under the European Community's Seventh Framework Programme to the OGLE project.
{\L}W acknowledges support from the ''Ideas Plus'' grant No.\ IdP2012 000162 by the Polish Ministry of Science and Higher Education.
SGD and AAM were supported in part by the NSF grant AST-0909182.

\begin{deluxetable*}{lcccccccccr}
\centering
\tabletypesize{\tiny}
\tablecaption{The SN observed magnitudes in the AB magnitude system.\label{tab:observed}}
\tablehead{HJD' (days) & $u$ & $\sigma_u$ & $g$ & $\sigma_g$ & $r$ & $\sigma_r$ & $i$ & $\sigma_i$ & $z$ & $\sigma_z$}
\startdata
3622.32727 & 20.588 & 0.093 & 20.728 & 0.065 & 20.745 & 0.055 & 20.904 & 0.086 & 22.295 & 1.424 \\
3634.27689 & 19.705 & 0.122 & 19.625 & 0.047 & 19.854 & 0.043 & 20.033 & 0.053 & 20.465 & 0.270 \\
3636.28243 & 19.527 & 0.073 & 19.575 & 0.028 & 19.766 & 0.036 & 19.960 & 0.042 & 20.632 & 0.372 \\
3638.31822 & 19.513 & 0.044 & 19.557 & 0.029 & 19.746 & 0.022 & 19.987 & 0.033 & 19.956 & 0.106 \\
3655.25412 & 19.634 & 0.095 & 19.320 & 0.028 & 19.523 & 0.032 & 19.709 & 0.051 & 19.774 & 0.154 \\
3666.24223 & 19.720 & 0.047 & 19.380 & 0.032 & 19.497 & 0.020 & 19.714 & 0.027 & 19.673 & 0.091 \\
3669.23771 & 19.835 & 0.047 & 19.433 & 0.024 & 19.549 & 0.021 & 19.737 & 0.026 & 19.930 & 0.104 \\
3671.32237 & 19.999 & 0.074 & 19.442 & 0.024 & 19.529 & 0.025 & 19.808 & 0.035 & 19.887 & 0.162 \\
3674.22166 & 20.009 & 0.058 & 19.550 & 0.020 & 19.561 & 0.024 & 19.795 & 0.031 & 19.764 & 0.102 \\
3676.31101 & 20.157 & 0.081 & 19.575 & 0.026 & 19.559 & 0.021 & 19.738 & 0.040 & 19.881 & 0.108 \\
3679.28270 & 20.081 & 0.074 & 19.545 & 0.025 & 19.551 & 0.027 & 19.858 & 0.037 & 20.172 & 0.196 \\
3681.27407 & 20.181 & 0.064 & 19.617 & 0.022 & 19.625 & 0.019 & 19.863 & 0.028 & 19.924 & 0.116 \\
3685.23284 & 20.338 & 0.200 & 19.786 & 0.067 & 19.632 & 0.063 & 19.929 & 0.066 & 20.266 & 0.265 \\
3694.24599 & 19.996 & 0.233 & 19.786 & 0.067 & 19.661 & 0.047 & 19.810 & 0.072 & 20.087 & 0.332 \\
3697.22983 & 20.948 & 0.146 & 20.011 & 0.027 & 19.844 & 0.027 & 20.022 & 0.030 & 19.981 & 0.131 \\
3700.22852 & 21.316 & 0.205 & 20.241 & 0.032 & 19.836 & 0.025 & 20.036 & 0.034 & 19.867 & 0.136 \\
3705.21473 & 22.056 & 0.333 & 20.389 & 0.030 & 19.990 & 0.030 & 20.052 & 0.034 & 20.007 & 0.127
\enddata
\tablecomments{$\rm HJD'=HJD - 2450000$, the magnitudes were corrected for the Milky Way's extinction and put on the AB magnitude system.}
\end{deluxetable*}

\begin{deluxetable*}{lcccccccccr}
\tabletypesize{\tiny}
\tablecaption{The SN absolute magnitudes in the AB magnitude system.\label{tab:absolute}}
\tablehead{HJD' (days) & $u$ & $\sigma_u$ & $g$ & $\sigma_g$ & $r$ & $\sigma_r$ & $i$ & $\sigma_i$ & $z$ & $\sigma_z$}
\startdata
3622.32727 & -20.182 & 0.093 & -20.042 & 0.065 & -20.025 & 0.055 & -19.866 & 0.086 & -18.475 & 1.424 \\
3634.27689 & -21.065 & 0.122 & -21.145 & 0.047 & -20.917 & 0.043 & -20.738 & 0.053 & -20.306 & 0.270 \\
3636.28243 & -21.243 & 0.073 & -21.195 & 0.028 & -21.004 & 0.036 & -20.810 & 0.042 & -20.138 & 0.372 \\
3638.31822 & -21.257 & 0.044 & -21.213 & 0.029 & -21.024 & 0.022 & -20.783 & 0.033 & -20.814 & 0.106 \\
3655.25412 & -21.136 & 0.095 & -21.450 & 0.028 & -21.247 & 0.032 & -21.061 & 0.051 & -20.996 & 0.154 \\
3666.24223 & -21.050 & 0.047 & -21.390 & 0.032 & -21.273 & 0.020 & -21.056 & 0.027 & -21.097 & 0.091 \\
3669.23771 & -20.935 & 0.047 & -21.337 & 0.024 & -21.221 & 0.021 & -21.033 & 0.026 & -20.840 & 0.104 \\
3671.32237 & -20.771 & 0.074 & -21.328 & 0.024 & -21.241 & 0.025 & -20.962 & 0.035 & -20.883 & 0.162 \\
3674.22166 & -20.761 & 0.058 & -21.220 & 0.020 & -21.209 & 0.024 & -20.975 & 0.031 & -21.006 & 0.102 \\
3676.31101 & -20.614 & 0.081 & -21.195 & 0.026 & -21.211 & 0.021 & -21.033 & 0.040 & -20.889 & 0.108 \\
3679.28270 & -20.689 & 0.074 & -21.225 & 0.025 & -21.219 & 0.027 & -20.912 & 0.037 & -20.598 & 0.196 \\
3681.27407 & -20.589 & 0.064 & -21.153 & 0.022 & -21.146 & 0.019 & -20.907 & 0.028 & -20.846 & 0.116 \\
3685.23284 & -20.432 & 0.200 & -20.984 & 0.067 & -21.138 & 0.063 & -20.841 & 0.066 & -20.504 & 0.265 \\
3694.24599 & -20.774 & 0.233 & -20.984 & 0.067 & -21.109 & 0.047 & -20.960 & 0.072 & -20.683 & 0.332 \\
3697.22983 & -19.822 & 0.146 & -20.759 & 0.027 & -20.926 & 0.027 & -20.748 & 0.030 & -20.789 & 0.131 \\
3700.22852 & -19.454 & 0.205 & -20.529 & 0.032 & -20.934 & 0.025 & -20.734 & 0.034 & -20.903 & 0.136 \\
3705.21473 & -18.714 & 0.333 & -20.381 & 0.030 & -20.780 & 0.030 & -20.718 & 0.034 & -20.763 & 0.127 
\enddata
\tablecomments{$\rm HJD'=HJD - 2450000$, the magnitudes were corrected for the Milky Way's extinction and put on the AB magnitude system, while using the distance modulus $\mu=40.77$~mag.}
\end{deluxetable*}

\begin{deluxetable*}{lcccccccccccr}
\tabletypesize{\tiny}
\tablecaption{Luminosity of the SN in units of $10^{43}$ erg/s.\label{tab:luminosity}}
\tablehead{HJD' (days) & $u$ & $\sigma_u$ & $g$ & $\sigma_g$ & $r$ & $\sigma_r$ & $i$ & $\sigma_i$ & $z$ & $\sigma_z$ & sum & $\sigma_{\rm sum}$ }
\startdata
3622.32727 & 4.193 & 0.376 & 2.898 & 0.180 & 2.170 & 0.114 & 1.544 & 0.128 & 0.359 & 0.974 & 11.164 & 1.073 \\
3634.27689 & 9.458 & 1.127 & 8.008 & 0.351 & 4.931 & 0.197 & 3.445 & 0.174 & 1.939 & 0.548 & 27.781 & 1.327 \\
3636.28243 & 11.140 & 0.776 & 8.385 & 0.218 & 5.344 & 0.180 & 3.684 & 0.145 & 1.661 & 0.678 & 30.214 & 1.078 \\
3638.31822 & 11.293 & 0.467 & 8.521 & 0.227 & 5.443 & 0.110 & 3.593 & 0.112 & 3.098 & 0.317 & 31.948 & 0.628 \\
3655.25412 & 10.095 & 0.922 & 10.600 & 0.274 & 6.683 & 0.199 & 4.642 & 0.224 & 3.661 & 0.559 & 35.681 & 1.152 \\
3666.24223 & 9.333 & 0.409 & 10.037 & 0.300 & 6.845 & 0.127 & 4.622 & 0.118 & 4.020 & 0.353 & 34.857 & 0.641 \\
3669.23771 & 8.389 & 0.374 & 9.554 & 0.216 & 6.528 & 0.129 & 4.522 & 0.111 & 3.171 & 0.320 & 32.164 & 0.563 \\
3671.32237 & 7.214 & 0.508 & 9.480 & 0.208 & 6.648 & 0.156 & 4.235 & 0.139 & 3.301 & 0.530 & 30.878 & 0.791 \\
3674.22166 & 7.148 & 0.393 & 8.576 & 0.161 & 6.455 & 0.144 & 4.287 & 0.125 & 3.697 & 0.363 & 30.163 & 0.590 \\
3676.31101 & 6.242 & 0.483 & 8.387 & 0.201 & 6.465 & 0.125 & 4.521 & 0.168 & 3.319 & 0.347 & 28.934 & 0.661 \\
3679.28270 & 6.690 & 0.474 & 8.622 & 0.200 & 6.515 & 0.164 & 4.046 & 0.140 & 2.539 & 0.501 & 28.412 & 0.749 \\
3681.27407 & 6.100 & 0.368 & 8.069 & 0.163 & 6.088 & 0.110 & 4.027 & 0.107 & 3.189 & 0.359 & 27.473 & 0.560 \\
3685.23284 & 5.280 & 1.070 & 6.901 & 0.440 & 6.046 & 0.360 & 3.791 & 0.238 & 2.328 & 0.642 & 24.346 & 1.391 \\
3694.24599 & 7.238 & 1.729 & 6.904 & 0.439 & 5.888 & 0.261 & 4.228 & 0.290 & 2.744 & 0.980 & 27.002 & 2.072 \\
3697.22983 & 3.011 & 0.434 & 5.612 & 0.140 & 4.974 & 0.126 & 3.480 & 0.099 & 3.028 & 0.390 & 20.105 & 0.621 \\
3700.22852 & 2.146 & 0.445 & 4.542 & 0.137 & 5.012 & 0.119 & 3.435 & 0.108 & 3.360 & 0.449 & 18.495 & 0.666 \\
3705.21473 & 1.085 & 0.389 & 3.960 & 0.111 & 4.347 & 0.123 & 3.385 & 0.106 & 2.954 & 0.366 & 15.731 & 0.569
\enddata
\tablecomments{$\rm HJD'=HJD - 2450000$, the luminosity is corrected for the Milky Way's extinction.}
\end{deluxetable*}




\begin{thebibliography}{}

\bibitem[Ahn et~al.(2012)]{2012ApJS..203...21A} Ahn, C.~P., et~al.\ 2012, \apjs, 203, 21A

\bibitem[Assef et al.(2010)]{2010ApJ...713..970A} Assef, R.~J., Kochanek, 
C.~S., Brodwin, M., et al.\ 2010, \apj, 713, 970 

\bibitem[Bramich et~al.(2008)]{2008MNRAS.386..887B} Bramich, D.~M., et~al.\ 2008, \mnras, 386, 887

\bibitem[Delahaye et al.(2010)]{2010arXiv1005.0423D} Delahaye, F., 
Pinsonneault, M.~H., Pinsonneault, L., 
\& Zeippen, C.~J.\ 2010, arXiv:1005.0423 

\bibitem[Drake et~al.(2009)]{2009ApJ...696..870D} Drake, A.~J., et~al.\ 2009, \apj, 696, 870

\bibitem[Elias-Rosa et~al.(2013)]{2013ATel.4774....1E} Elias-Rosa, N., et~al.\ 2013, ATel, 4774, 1

\bibitem[Filippenko(1997)]{1997ARA&A..35..309F} Filippenko, A.~V.\ 1997, \araa, 35, 309 

\bibitem[Frieman et~al.(2008)]{2008AJ....135..338F} Frieman, J.~A., et~al.\ 2008, \aj, 135, 338

\bibitem[Gal-Yam et~al.(2009)]{2009Natur.462..624G} Gal-Yam, A. et~al.\ 2009, \nat, 462, 624

\bibitem[Gal-Yam(2012)]{2012Sci...337..927G} Gal-Yam, A.\ 2012, Science, 337, 927

\bibitem[Inserra et~al.(2013)]{2013arXiv1304.3320I} Inserra, C., et~al.\ 2013, ArXiv, 1304.3320

\bibitem[Ivezi{\'c} et al.(2007)]{2007AJ....134..973I} Ivezi{\'c}, {\v Z}., 
Smith, J.~A., Miknaitis, G., et al.\ 2007, \aj, 134, 973 

\bibitem[Kim et al.(1996)]{1996PASP..108..190K} Kim, A., Goobar, A., 
\& Perlmutter, S.\ 1996, \pasp, 108, 190 

\bibitem[Koz{\l}owski et al.(2010)]{2010ApJ...722.1624K} Koz{\l}owski, S., 
Kochanek, C.~S., Stern, D., et al.\ 2010, \apj, 722, 1624 

\bibitem[Koz{\l}owski et al.(2013)]{2013AcA....63....1K} Koz{\l}owski, S., 
Udalski, A., Wyrzykowski, {\L}., et al.\ 2013, Acta Astron., 63, 1

\bibitem[Law et~al.(2009)]{2009PASP..121.1395L} Law, N.~M., et~al.\ 2009, \pasp, 121, 1395

\bibitem[Miller et~al.(2009)]{2009ApJ...690.1303M} Miller, A.~A., et~al.\ 2009, \apj, 690, 1303

\bibitem[Modjaz et~al.(2009)]{2009ApJ...702..226M} Modjaz, M., et~al.\ 2009, \apj, 702, 226

\bibitem[Neill et~al.(2011)]{2011ApJ...727...15N} Neill, J.~D. et~al.\ 2011, \apj, 727, 15

\bibitem[Pastorello et~al.(2010)]{2010ApJ...724L..16P} Pastorello, A., et~al.\ 2010, \apjl, 724, L16

\bibitem[Pettini \& Pagel(2004)]{2004MNRAS.348L..59P} Pettini, M., \& Pagel, B.~E.~J.\ 2004, \mnras, 348, L59 

\bibitem[Prieto et al.(2008)]{2008ApJ...673..999P} Prieto, J.~L., Stanek, 
K.~Z., \& Beacom, J.~F.\ 2008, \apj, 673, 999 

\bibitem[Quimby et~al.(2011)]{2011Natur.474..487Q} Quimby, R.~M., et~al.\ 2011, \nat, 474, 487

\bibitem[Rau et~al.(2009)]{2009PASP..121.1334R} Rau, A., et~al.\ 2009, \pasp, 121, 1334

\bibitem[Sako et~al.(2008)]{2008AJ....135..348S} Sako, M., et~al.\ 2008, \aj, 135, 248

\bibitem[Schlegel et al.(1998)]{1998ApJ...500..525S} Schlegel, D.~J., 
Finkbeiner, D.~P., \& Davis, M.\ 1998, \apj, 500, 525

\bibitem[Stanek et al.(2006)]{2006AcA....56..333S} Stanek, K.~Z., Gnedin, 
O.~Y., Beacom, J.~F., et al.\ 2006, Acta Astron., 56, 333 

\bibitem[Stoll et al.(2011)]{2011ApJ...730...34S} Stoll, R., Prieto, J.~L., 
Stanek, K.~Z., et al.\ 2011, \apj, 730, 34 

\bibitem[Tremonti et al.(2004)]{2004ApJ...613..898T} Tremonti, C.~A., 
Heckman, T.~M., Kauffmann, G., et al.\ 2004, \apj, 613, 898 

\bibitem[Young et~al.(2010)]{2010A&A...512A..70Y} Young, D.~R., et~al.\ 2010, \aap, 512, A70

\bibitem[Wyrzykowski et~al.(2012)]{2012ATel.4495....1W} Wyrzykowski, L., Udalski, A., \& Kozlowski, S.\ 2012, ATel, 4495, 1


\end{thebibliography}
\end{document}